\documentclass[a4paper,10pt,twoside]{cpc-hepnp}
\usepackage{CJK,upgreek,fancyhdr}
\usepackage{multicol}
\usepackage{graphicx}
\usepackage{booktabs}
\usepackage{amssymb,bm,mathrsfs,bbm,amscd}
\usepackage[tbtags]{amsmath}
\usepackage{lastpage}
\usepackage{color}

\begin{document}
\begin{CJK*}{GB}{gbsn}

\fancyhead[c]{\small XXXXXX~~~Vol. xx, No. x (2017) xxxxxx}
\fancyfoot[C]{\small 010201-\thepage}

\footnotetext[0]{Received xx February 2017}

\title{Measurements of the Total Cross section for Thermal Neutrons at PNS\thanks{Supported by the Chinese
TMSR Strategic Pioneer Science and Technology Project (No.XDA02010100),
 National Natural Science Foundation of China (NSFC)(No.11475245,No.11305239)}}

\author{%
      Longxiang Liu $^{1;1)}$\email{liulongxiang@sinap.ac.cn},%
\quad Hongwei Wang $^{1;2)}$\email{wanghongwei@sinap.ac.cn},%
\quad Yugang Ma $^{1}$,
\quad Xiguang Cao $^{1}$,
\quad Xiangzhou Cai $^{1}$,\\
\quad Jingen Chen $^{1}$,
\quad Guilin Zhang $^{1}$,
\quad Jianlong Han $^{1;3)}$\email{hanjianlong@sinap.ac.cn},%
\quad Guoqiang Zhang $^{1}$,
\quad Jifeng Hu $^{1}$,\\
\quad Xiaohe Wang $^{1,2}$,
\quad Wenjiang Li $^{1}$,
\quad Zhe Yan $^{1}$,
and \quad Haijuan Fu $^{1,2,3}$
}
\maketitle

\address{%
$^1$ Shanghai Institute of Applied Physics, Chinese Academy of Sciences, Shanghai 201800, China\\
$^2$ University of Chinese Academy of Science, Beijing 100080, China\\
$^3$ Shanghai University, School of Materials Science and Engineering, Shanghai 200444, China\\
}

\begin{abstract}
In order to measure the total cross section for thermal neutrons, a photoneutron source (PNS, phase 1)
has been developed for the acquisition of nuclear data for the Thorium Molten Salt Reactor (TMSR)
at the Shanghai Institute of Applied Physics (SINAP).
PNS is an electron LINAC pulsed neutron facility that uses the time-of-flight (TOF) technique.
It records the neutron TOF and identifies neutrons and $\gamma$-rays by using a digital signal processing technique.
The background is obtained by using a combination of employing 12.8 cm boron-loaded polyethylene(PEB) (5$\%$ w.t.)
to block the flight path and  Monte Carlo methods.
The neutron total cross sections of natural beryllium are measured in the neutron energy region from 0.007 to 0.1 eV.
The present measurement result is compared with the fold Harvey data with the response function of PNS.
\end{abstract}



\section{Introduction}

Neutron total cross sections have been investigated for over 70 years using a transmission method.
With the development of scientific technology, the measurement method has been improved and the energy
range of neutrons has been expanded.

 In 1941, Hanstein used the resonance filter method to define the energy of a neutron beam from a paraffin block,
 which was utilized to slow down fast neutrons produced by the proton-beryllium reaction in the cyclotron.
 Its measuring system consisted of an uranium oxide ionization chamber connected to an Edelman string electrometer \cite{HENRY}.
 The disadvantage of the resonance filter method is that only a few points of neutron energy could be measured.
 In 1947, Brill and Lichtenberger used the rotating shutter mechanism  for the velocity selection of slow neutrons from the
 thermal column of the Argonne heavy-water pile \cite{Brill}.
  The rotating shutter mechanism was the prototype of the neutron energy measurement system using the time-of-flight (TOF) method,
  and the energy range of neutrons was expanded significantly.
 There is a similar experimental device at the heavy water thermal neuron facility of the Kyoto University Research Reactor (KUR).
 A $^{3}$He counter was used to detect chopped neutrons from 0.001 to 0.025 eV \cite{Keiji}.

The pulsed-beam TOF technique was developed at the Weapons Neutron Research facility at the Los Alamos National Laboratory (LANL).
  Neutrons were produced by the spallation of an 800-MeV proton beam incident on a tungsten target
  and detected using a BC404 detector \cite {Finlay}.
  The neutron total cross sections are determined with a continuous spectrum of neutron energies to 600 MeV.
  Another neutron TOF facility (n\underline{ }TOF) at CERN allows high-resolution measurements of the neutron-induced cross sections with
  a 20 GeV/c  proton beam, pure lead target, and 185-m neutron flight length \cite{Borcea}.

  The TOF-facility GELINA consists of an 80-140 MeV electron linear accelerator, rotary mercury-cooled uranium target,
  post-acceleration relativistic-energy compression magnet system,
  12 different flight paths ranging from 10 m up to 400 m, and Li-glass, plastic scintillators or NE213 scintillators as a neutron detector.
  The neutron total cross sections were measured in the neutron energy region from 1 meV to 20 MeV \cite{Mondelaers}.
  The Gaerttner Laboratory's 55-MeV electron linear accelerator at Rensselaer Polytechnic Institute was used in the measurement of the neutron
  total cross sections in the energy range from 0.4 to 20 MeV with a 100 m flight path, fast detector response and electronics,
   and narrow  pulse width to provide a good energy resolution.
  A method to determine the time-dependent background component was implemented
  using a combination of experimental data and Monte Carlo methods \cite{Rapp}.
  In 2010, a digital-signal-processing method, which allowed a high statistical accuracy, was introduced to
   measure the total neutron cross sections \cite{Shane}.
  There are some other electron pulsed neutron sources, such as ORELA\cite{ORELA}, POHANG\cite{Taofeng}, and nELBE\cite{ELBE}.

In the present work, the system for measuring the neutron total cross sections consists of a 16 MeV electron accelerator,
 water-cooled tungsten target (diameter is 60 mm and length is 48 mm) with a 10-cm-long polyethylene moderator,
 and 6.2-m-long flight path \cite{jiangmianheng}.
 As a result of the limitation of experimental space, the tungsten target chamber and the neutron detector were shielded to
 reduce the background.
  A method utilizing 12.8 cm boron polyethylene (PEB) to block the neutron flight path and Monte Carlo was used to determine
   the time-dependent background component.
A digital-signal-processing technique was used in the data acquisition of the system.

\section{Experimental Setup}

The PNS is a compact-type system for measuring the neutron total cross sections, and all its devices are arranged in the experimental
hall in a 11 m$\times$8 m space, as shown in Fig.~\ref{PNF}.
Although it has the advantage of a small space, it has the disadvantage of a high background.
In order to reduce the background, the tungsten target chamber is shield by 5 cm Al, 25 cm Pb, 15 cm polyethylene (PE), and
5 cm Al, in sequence.
 At the entrance of the electron beam, it is shield by 10 cm Al, 10 cm Pb, and 10 cm PEB.
There is an L-type wall near the neutron detector, which consists of 30 cm concrete and 30 cm PEB.
The TOF detector is shielded by 5 cm Fe and 30 cm PEB, and the monitor detector by 20 cm PEB.
\begin{center}
\includegraphics[width=14cm]{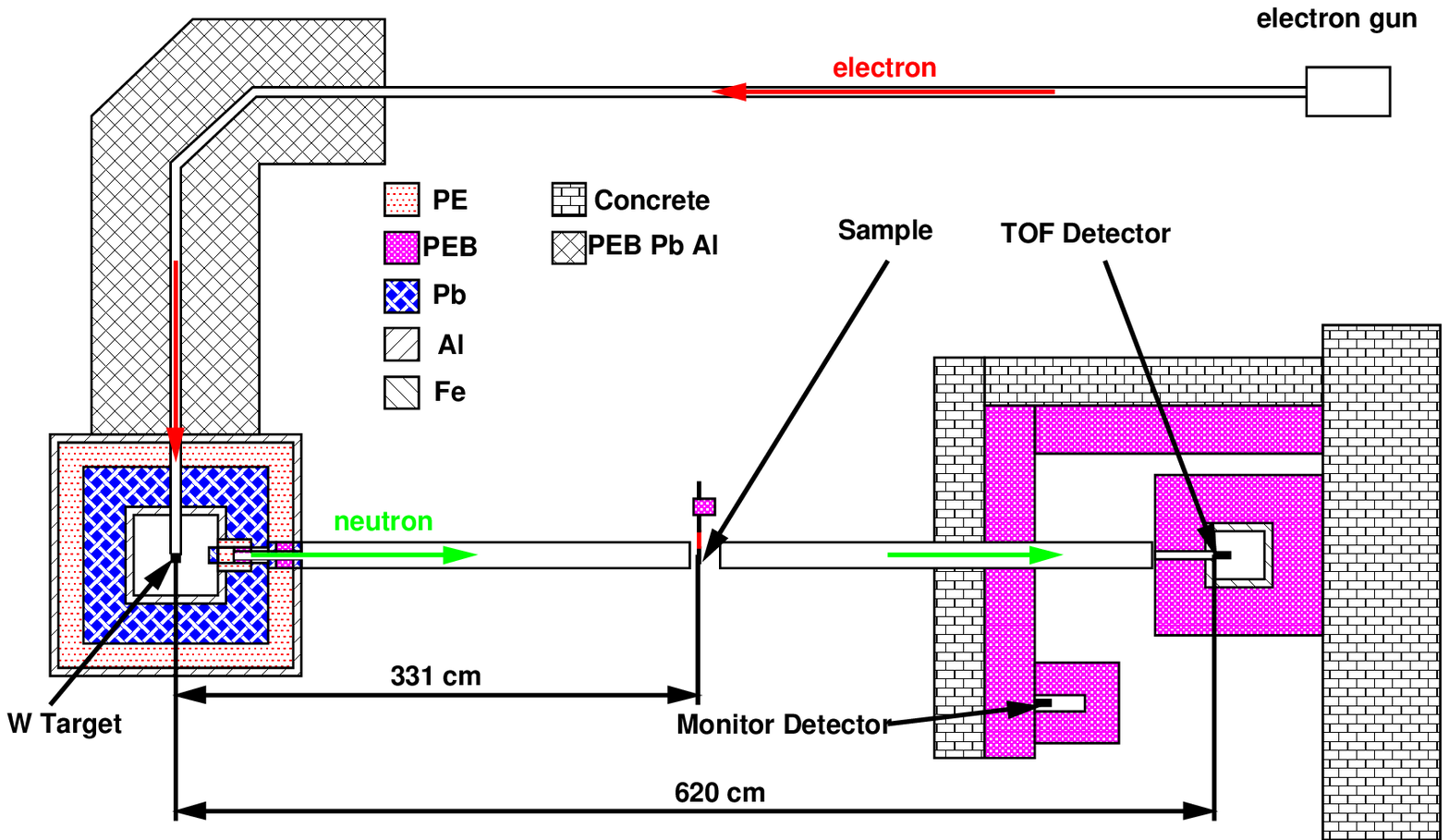}
\figcaption{\label{PNF}    Experimental geometry (plan view) of the PNS. }
\end{center}

Electrons are stopped in a heavy target by the bremsstrahlung mechanism,
and the radiation pulse produced in the detector is called the $\gamma$ flash.
 At the same time, neutrons are produced by the ($\gamma$,n) reaction with the target element.
The neutrons and $\gamma$ from the target go through a Pb plate with a thickness of 5 cm and a diameter of 10 cm
to reduce the $\gamma$ flash.
To maximize the thermal neutrons in this facility, we utilize a PE plate with a thickness of 10 cm and a diameter of 20 cm .
A pulsed neutron beam was collimated to a diameter of 5 cm by  a 10 cm PEB tube, 15 cm Pb tube,
 10 cm PEB tube, and 5 cm Pb tube, in sequence, as shown in Fig.~\ref{target}.

The sample changer was located at 331 cm from the target.
Five samples (including an open target) were held in the sample changer which cycled at set intervals, typically waiting 300 s on
each sample.
The first sample was a natural beryllium ($^{9}$B) plate with a thickness of 10 mm, and $10\times10$ cm$^{2}$ in the cross-sectional area.
 The second sample was high-purity (99.995$\%$) indium with a thickness of 0.1 mm
 (the data for this one is not discussed here, because it is irrelevant to this paper).
The third sample was a set of notch filters of Cd (purity 99.99$\%$) with 0.125-mm-thickness, Co (purity 99.9$\%$) with 0.05-mm-thickness,
Ag (purity 99.95$\%$) with 0.1-mm-thickness, and In (purity 99.99$\%$) with 0.05-mm-thickness plates.
This sample was used for energy calibration.
The fourth sample was a PEB plate with a thickness of 12.8 cm for the background measurement.
 The last one was a blank sample.
 They all had the same size as that of the Be sample in the cross-sectional area.
In the present experiment, the data were collected for 40 min for each sample.

With regard to neutron detector, we used a $^6$LiF(ZnS) scintillator, with the product code of EJ426HD2 from Eljen company, a diameter of 50 mm, and
a thickness of 0.5 mm, mounted on an EMI9813 photomultiplier produced by ET enterprise Ltd \cite{EJ,dulong,chang1,chang2}.
There were two identical neutron detectors, called TOF detector and monitor detector.
Transmitted neutrons were detected 620 cm downstream of the production target in the TOF detector.
The monitor one was placed at the bottom to the right of the neutron beam pipe, and the data from this monitor were
used to correct any variations in beam intensity.

\section{ Data acquisition}

The neutron energy spectra produced from the tungsten target with a 10-cm-long polyethylene moderator were obtained by using the TOF technique.
We used a CAEN DT5720 digitizer, four-channel 12 bit 250 MS/s waveform digitizer (WFD) with a bandwidth of 125 MHz,
 and 2 Vpp dynamic ranges on single-ended coax MCX input connectors in this work.
 The acquisition continued into a new buffer, without a dead time \cite{caen}.
 Therefore, without signal overlapping, there is no need to consider dead time correction.
In this work, waveform traces with 512 samples  were stored for the off-line analysis,
and the post-trigger phase waveform comprised 80$\%$ of the entire acquisition window.

The signal obtained by the TOF detector was connected to channel 1 of the digitizer, and the signal for
 monitoring the beam intensity was connected to channel 2.
The trigger signal (the electron GUN start signal) was connected to channel 3,
used as the start signal for the TOF calculation.
 The digitizer was connected to a personal computer with a Linux operating system via an optical cable.
 Because of different baselines, the thresholds of the three channels were set as -14.65, -4.88, and -4.88 mV.
  The self-trigger mode was used,
 which means that any channel signal above the threshold will trigger the DAQ.
 The trigger occurs on the falling edge of the signal.
The data acquisition software known as TMSR-Digitizer-DAQ was developed based on ROOT \cite{root} and Gnuplot \cite{gnuplot}.
The digitizer houses the optical link interface, which supports a transfer rate of 80 MBytes/s.
It is a very high-efficiency DAQ system for millisecond- or second-level neutron TOF spectroscopy, which incorporates data compression,
automatic change of the target, automatic alarm of beam loss, and automatic file opening functions \cite{Llx}.

During the experiment, the electron accelerator was operated with a repetition rate of 40 Hz, a pulse
width of 1 $\mu$s, a peak current of 10 $\mu$A, and an electron energy of 16 MeV.

\section{Simulation Method}

In order to obtain the response function and the time-dependent background component of this system,
a Monte Carlo simulation code based on Geant4 was implemented.
The version of Geant4 was Geant4.10.3 with G4NDL4.5, neutron data files with thermal cross sections.

Considering the calculation time, the local weighted method was used in the simulation.
At the first step, particles with information on time, position, direction, energy, and type at position 1 (as shown in Fig.~\ref{target})
were recorded in the simulation of the electrons from an electron tube with an energy of 16 MeV bombarding the tungsten target.
At the second step, the particles recorded at position 1 were increased in number to continue the simulation.
The number of neutrons detected by the TOF detector was approximately 14 in the simulation of electrons with 16 MeV, $10^{13}$.
It was approximately 39 with the same number of electrons in the experiment.
From the comparison of these results, it was concluded that the physical list used in the simulation was reliable
despite the error of the simulation being large due to low statistics.

In order to further reduce the computation time, the neutrons with information on time and energy at position 2
(as shown in Fig.~\ref{target}) were recorded from the second step of the simulation.
Then, the positions of the neutron source were chosen randomly at position 2,
and the directions were all set to be parallel to the TOF tube at the third step of the simulation.
The five different samples were all simulated as the experimental process, and the results of that were compared with the ones
of the experiment, which are discussed in the next section.

In the simulation, we considered almost all factors influencing the measurements, such as the pulse width (1 $\mu$s),
 slowing time of PE with a thickness of 10 cm, time precision of electronics, and geometrical factors.

\begin{center}
\includegraphics[width=8cm]{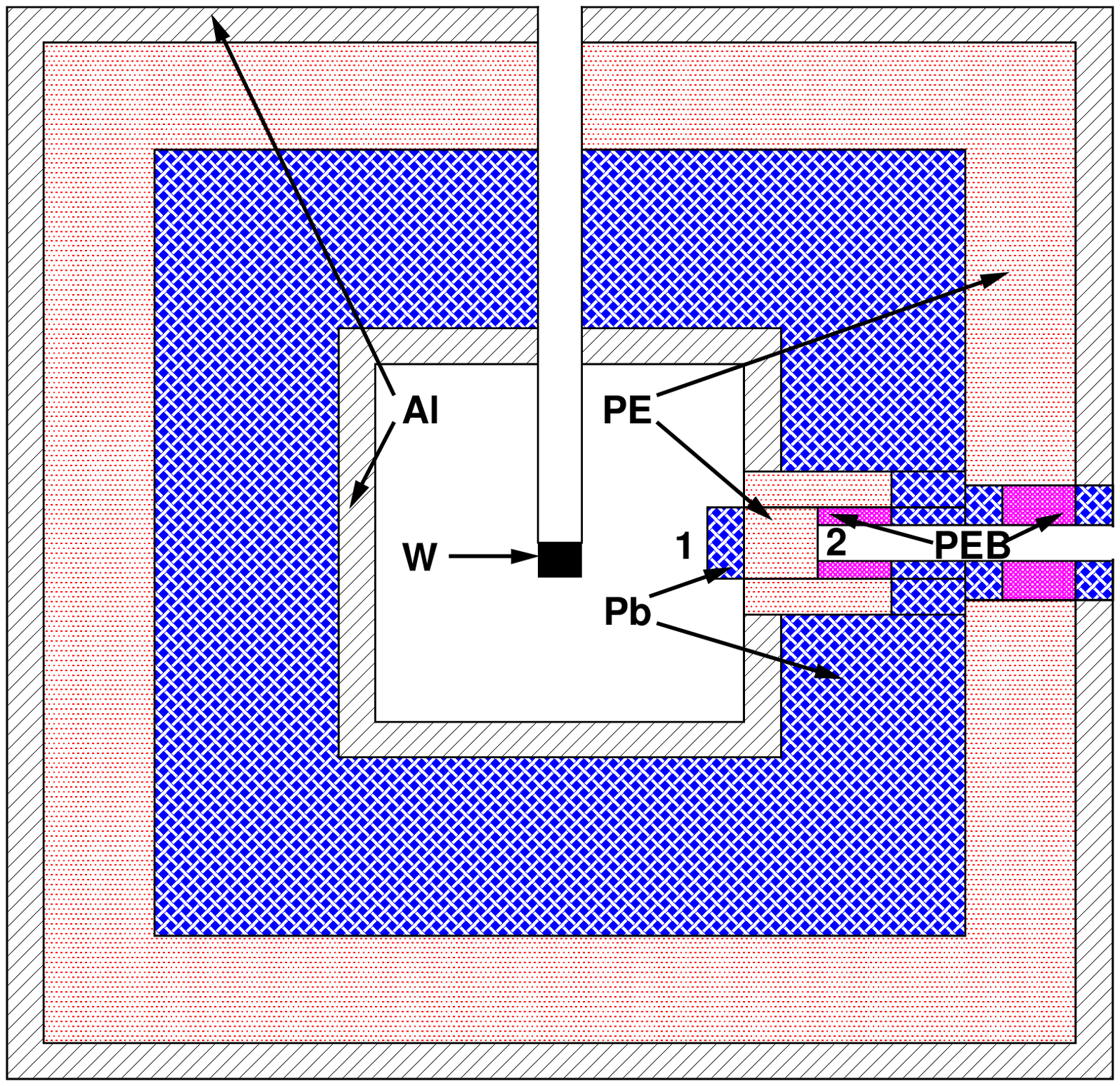}
\figcaption{\label{target}    Geometry (plan view) of the tungsten target chamber in the simulation. Position 1 is the left surface of
 the Pb plate with a diameter of 10 cm; Position 2 is the left surface of the neutron tube with a diameter of 5 cm. }
\end{center}

\section{Data Analysis and Results}

The TOF spectrum was only measured in the direction perpendicular to the incident electron beam.
 The method applied for the n/$\gamma$ identification for the TOF spectrum calculations was the pulse-shape discrimination
 (PSD, calculated by using the integral lengths) method \cite{chang1,Llx}.
 The falling edge of the electron GUN RF signal was used as the start signal, and the neutron peak position of the TOF detector
was used as the stop signal for the neutron TOF measurement.

\subsection{Response of a Time-of-Flight Spectrometer by Simulation}
The response function of a TOF-spectrometer $R(t_{m},E)$ is the probability that a neutron  with energy $E$ is detected
with a time-of-flight $t_{m}$.
It can be considered as a convolution of different independent components \cite{Schillebeeckx}.
In the case of the PNS facility the broadening in time is dominated by the neutron transport in the target-moderator assembly
($t_{t}$).
Consequently, the response functions will strongly depend on the neutron physics properties of the target-moderator assembly
(dimensions and materials).

\begin{center}
\includegraphics[width=8cm]{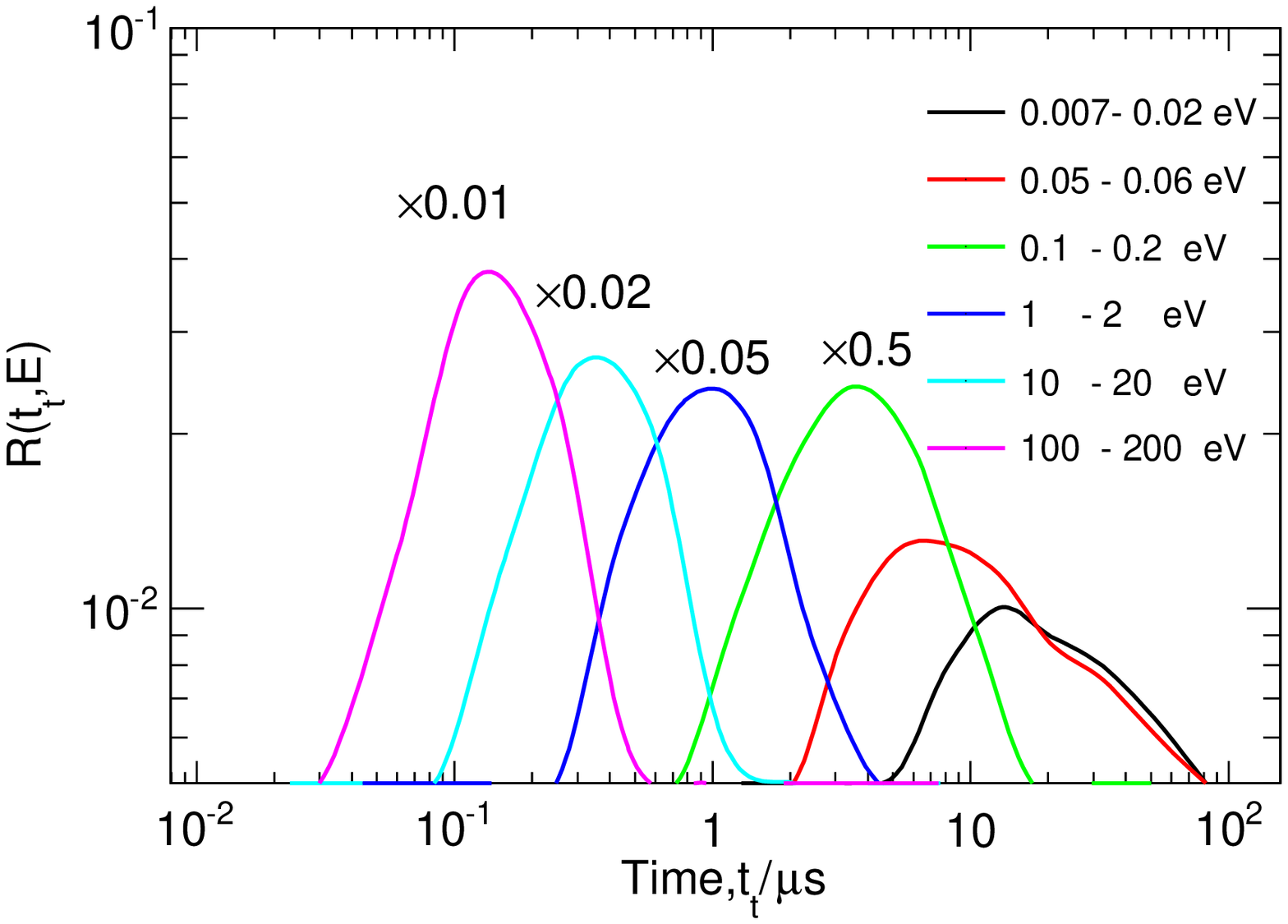}
\figcaption{\label{RTEM1}    The probability distribution of the time $t_{t}$ that a neutron spends in the target-moderator assembly of PNS. }
\end{center}

Monte Carlo (MC) simulations to determine the probability distribution of the time the neutron spends in the
target-moderator assembly have been carried out for the PNS facility.
Fig.~\ref{RTEM1} shows the response functions due to the neutron transport in the target-moderator assembly for the PNS facility.
The distributions are for a moderated neutron beam and a flight path that  forms an angle of $0^{\circ}$ with
the normal to the exit face of the moderator.
These distributions strongly depend on the neutron energy.
\begin{multicols}{2}

\begin{center}
\includegraphics[width=8cm]{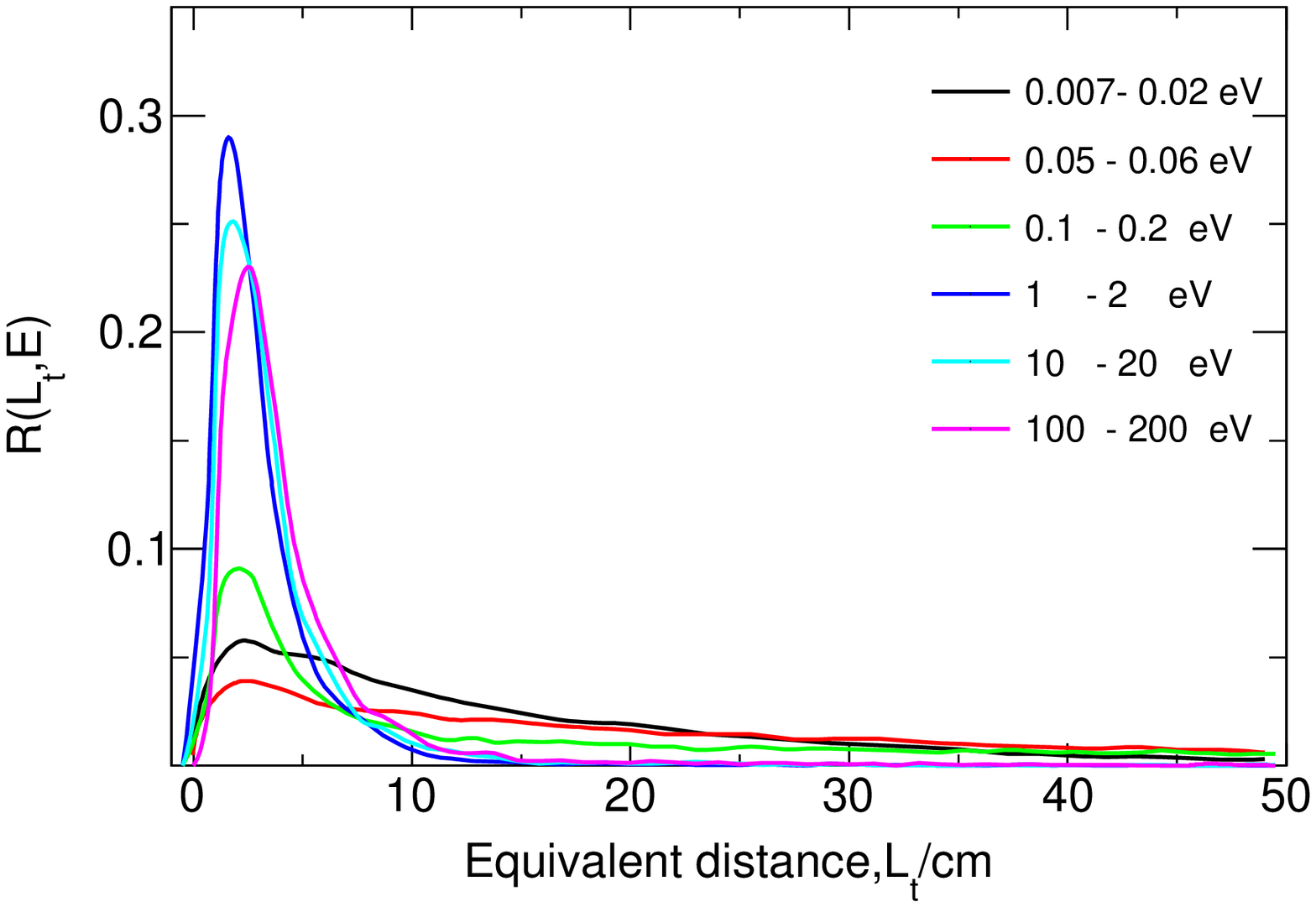}
\figcaption{\label{RLEM1}    The probability distribution of the equivalent distance $L_{t}$ that a neutron travels in the
 target-moderator assembly of PNS. }
\end{center}

\begin{center}
\includegraphics[width=8cm]{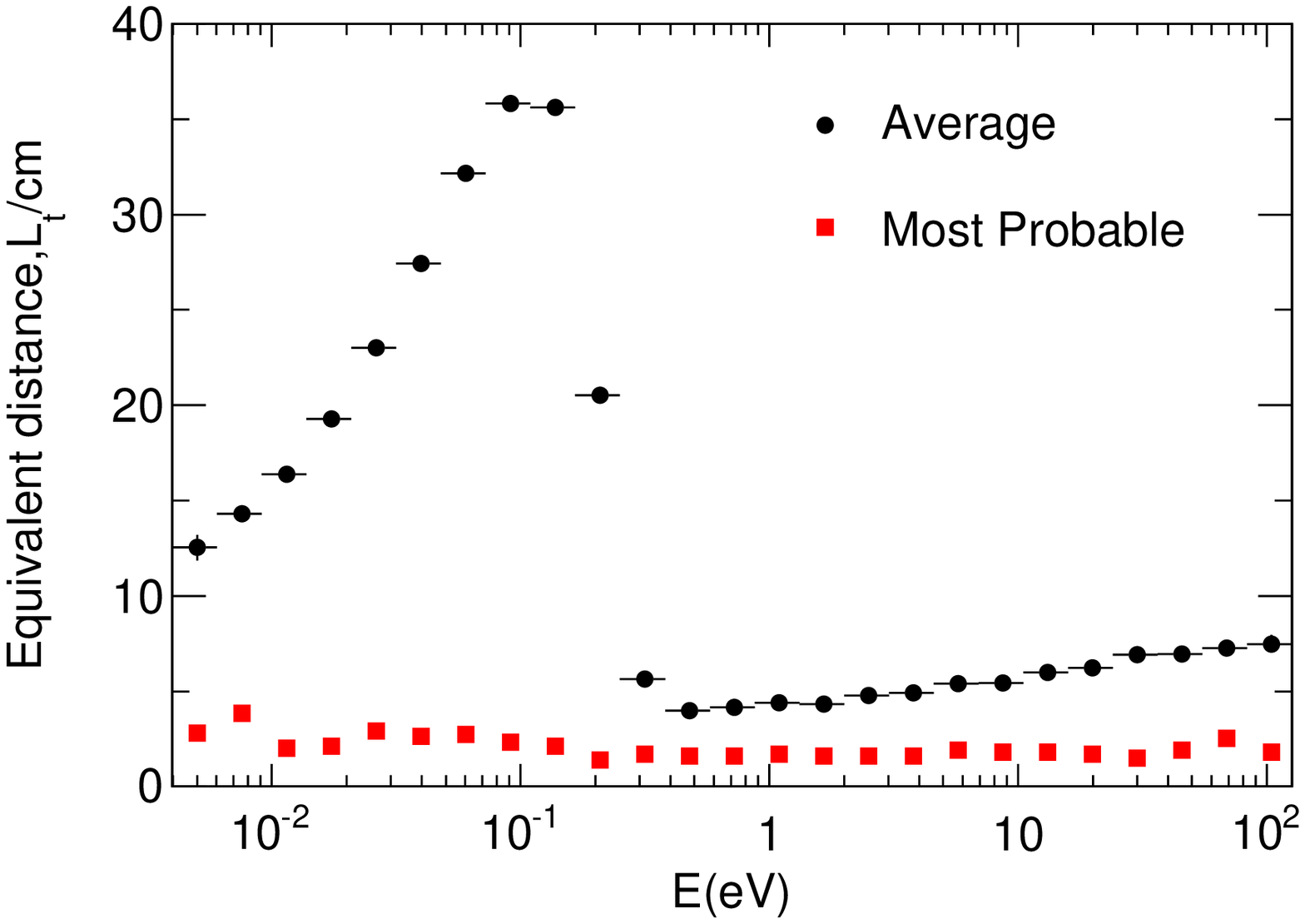}
\figcaption{\label{AverageL}    The Average and most probable equivalent distance as a function  of neutron energy for the target-moderator
 assembly of PNS. }
\end{center}
\end{multicols}

Response functions of a TOF-spectrometer can be more conveniently represented by introducing an equivalent distance $L_{t}$
travelled by the neutron in the target-moderator assembly.
The equivalent distance is defined as $L_{t}=vt_{t}$, where $v$ is the velocity of the neutron at the moment it escapes from
the target-moderator assembly.
This transformation of variables results in probability distributions of $L_{t}$ which are shown in Fig.~\ref{RLEM1}.
The energy dependence of the average equivalent distance and most probable distance is shown in Fig.~\ref{AverageL}.
The energy resolutions with the contribution due to the target-moderator assembly are shown in Fig.~\ref{energyError}.

\begin{center}
\includegraphics[width=8cm]{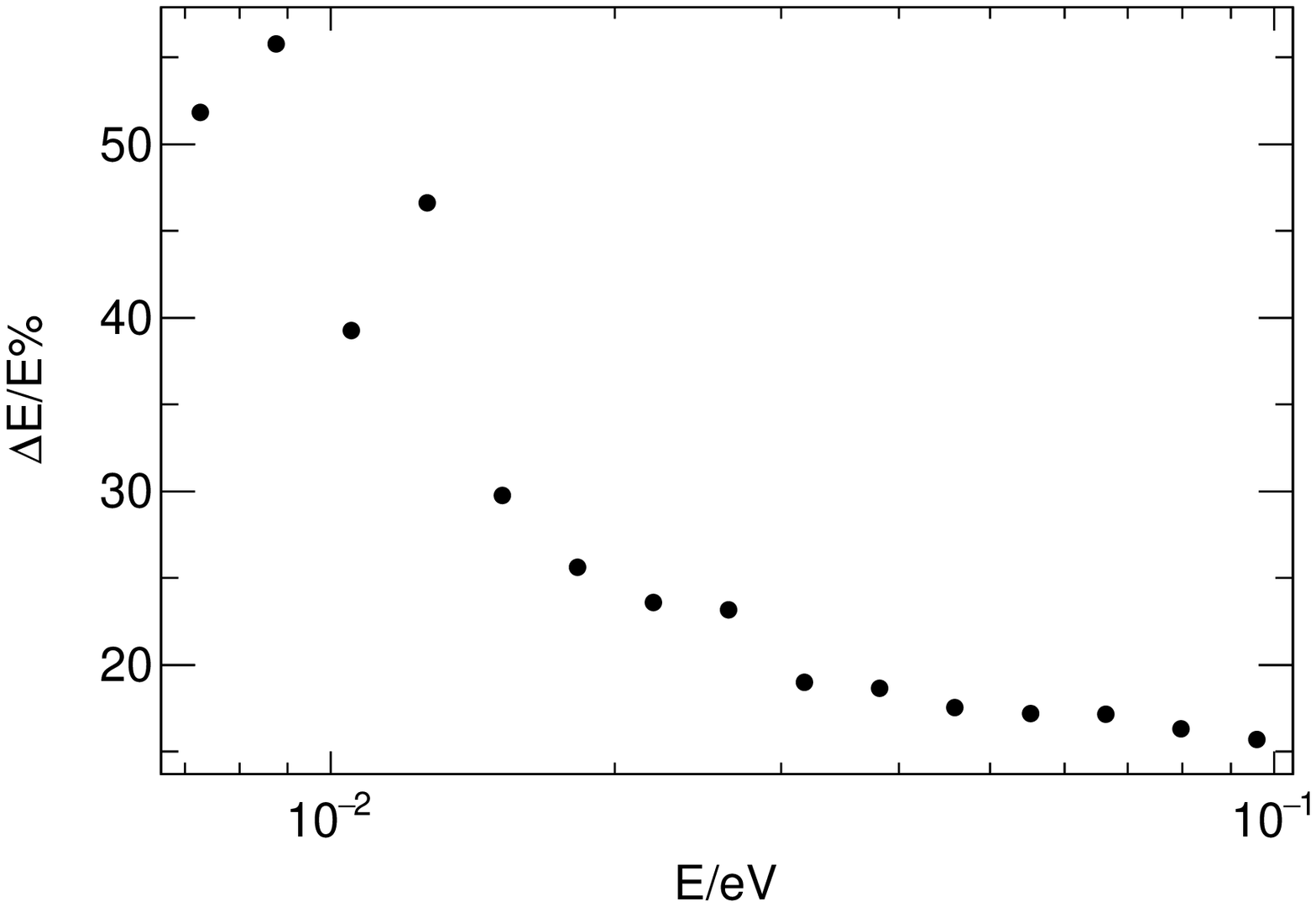}
\figcaption{\label{energyError}    The energy resolutions (at root-mean-square) for transmission measurements at PNS
 with the contribution due to the target-moderator assembly in the neutron energy range from 0.007 to 0.1 eV. }
\end{center}

\subsection{Background}

The background in the TOF transmission measurement at PNS can be considered as a sum of a time independent and time dependent
components
\begin{equation}\label{background}
  B(t)=B_{0}+B_{no}(t)+B_{ns}(t)+B_{ne}(t).
\end{equation}
The first time dependent component $B_{no}(t)$ results from overlap neutrons, \emph{i.e.} neutrons which
 are detected but have been produced in a previous cycle.
 The component $B_{no}(t)$, which strongly depends on the operating frequency,
 can be neglected at a low operating frequency of the accelerator.
 A second time dependent component $B_{ns}(t)$ originates predominantly from beam neutrons which are
 scattered inside the detector station .
 The $B_{ns}(t)$ of the open target and Be sample are shown in Fig.~\ref{bgMC} by simulation.
  A third time dependent component $B_{ne}(t)$ is due to neutrons scattered in the surroundings (environment).
 The time dependence background is hard to be distinguished from the time
 independent background $B_{0}$.
 The overall contribution $B_{0}+B_{ne}(t)$ can be determined by measurements with the beam closed.

\begin{center}
\includegraphics[width=12cm]{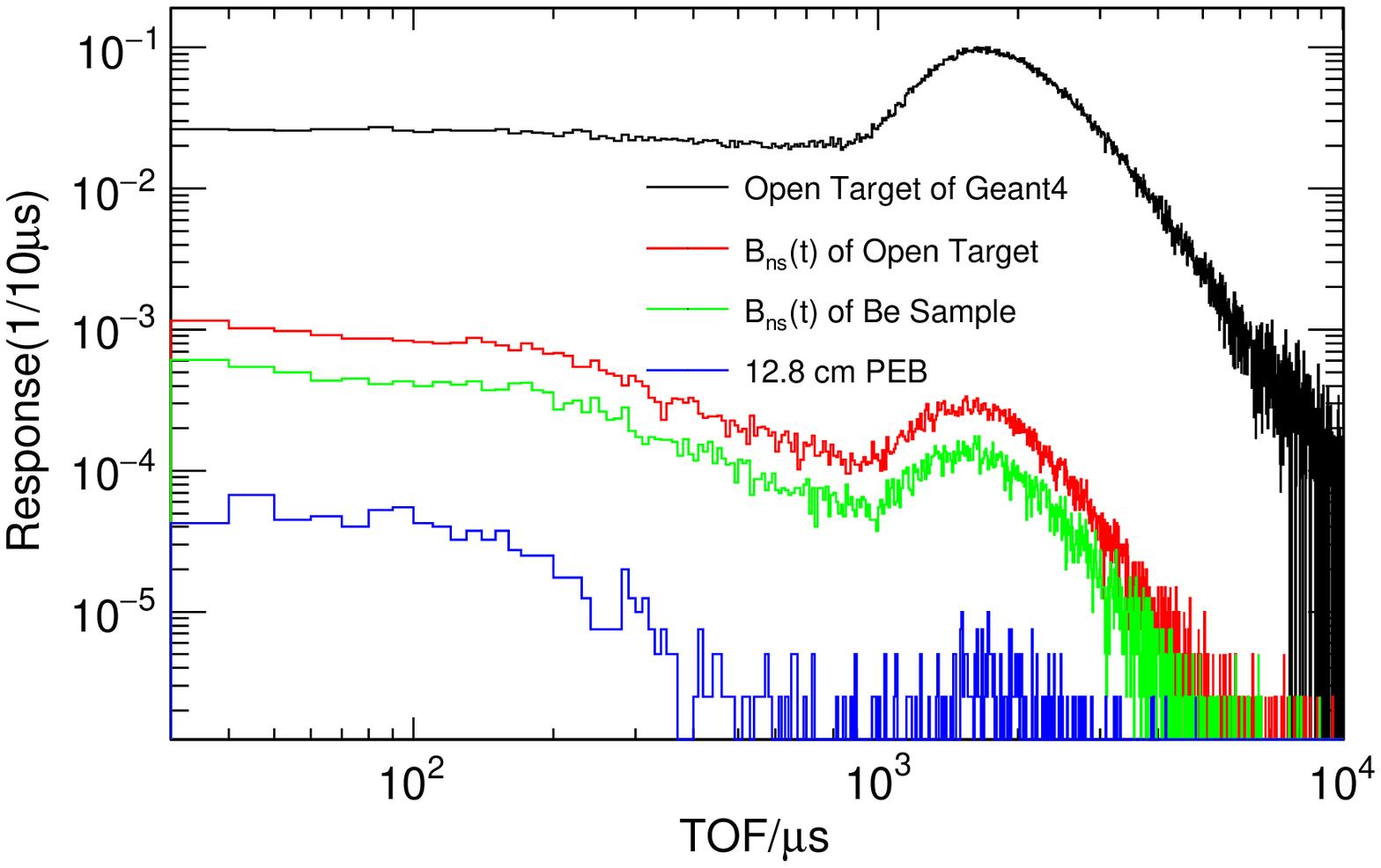}
\figcaption{\label{bgMC}   The response of a EJ426 neutron detector as a function of TOF by simulation,
and the $B_{ns}(t)$ background of the open target, Be sample, and the TOF spectrum of the
 neutrons transmitted through 12.8 cm PEB. }
\end{center}

The $B_{0}+B_{ne}(t)$ background level was estimated by using the background sample with 12.8-cm-thick PEB.
The TOF spectrum of the neutrons transmitted through PEB is shown in Fig.~\ref{bgMC}.
This proved that the $B_{0}+B_{ne}(t)$ background level in Fig.~\ref{tof} is reliable.
It was also proved to be accurate by the neutron TOF spectrum for the notch filter of 0.05 mm Co, 0.1 mm Ag, 0.05 mm
In, and 0.125 mm Cd samples in Fig.~\ref{tof}.
The dip of Co was not enough deep because of the transmission rate and $B_{ns}(t)$, as shown in Fig.~\ref{bgMC}.
The transmission rates of Co, Ag, and In were approximately 0.009, $3\times10^{-7}$, and $8\times10^{-6}$, respectively.

\begin{center}
\includegraphics[width=12cm]{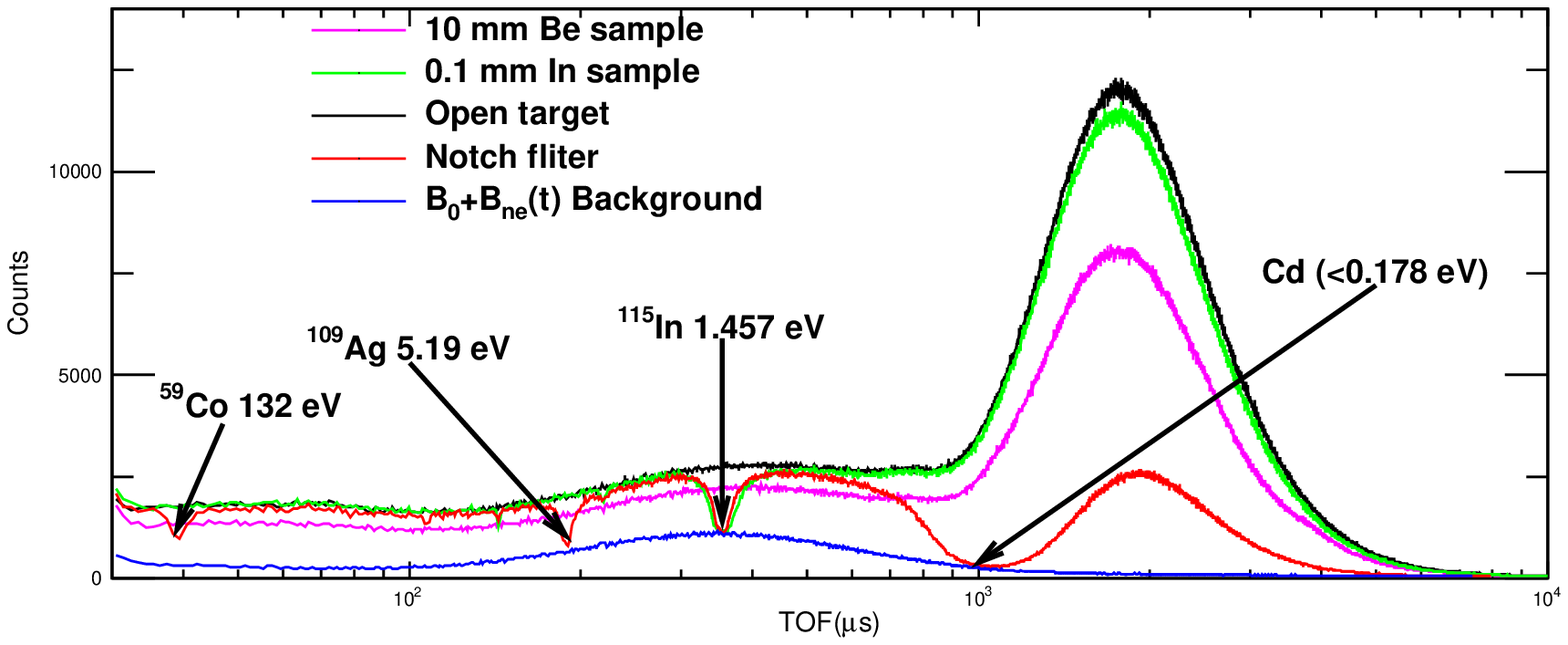}
\figcaption{\label{tof}    Experimental TOF spectra for the Be sample, In sample, open target, notch filter,
and $B_{0}+B_{ne}(t)$ background. }
\end{center}

In Fig.~\ref{tofME}, the TOF spectra for the open target and notch filter of the experiment have subtracted background.
The TOF spectra for the open target and notch filter of the experiment were consistent with the ones obtained
by the simulation in the time and shape, except for some differences in the number of counts in the lower time region.

\begin{center}
\includegraphics[width=12cm]{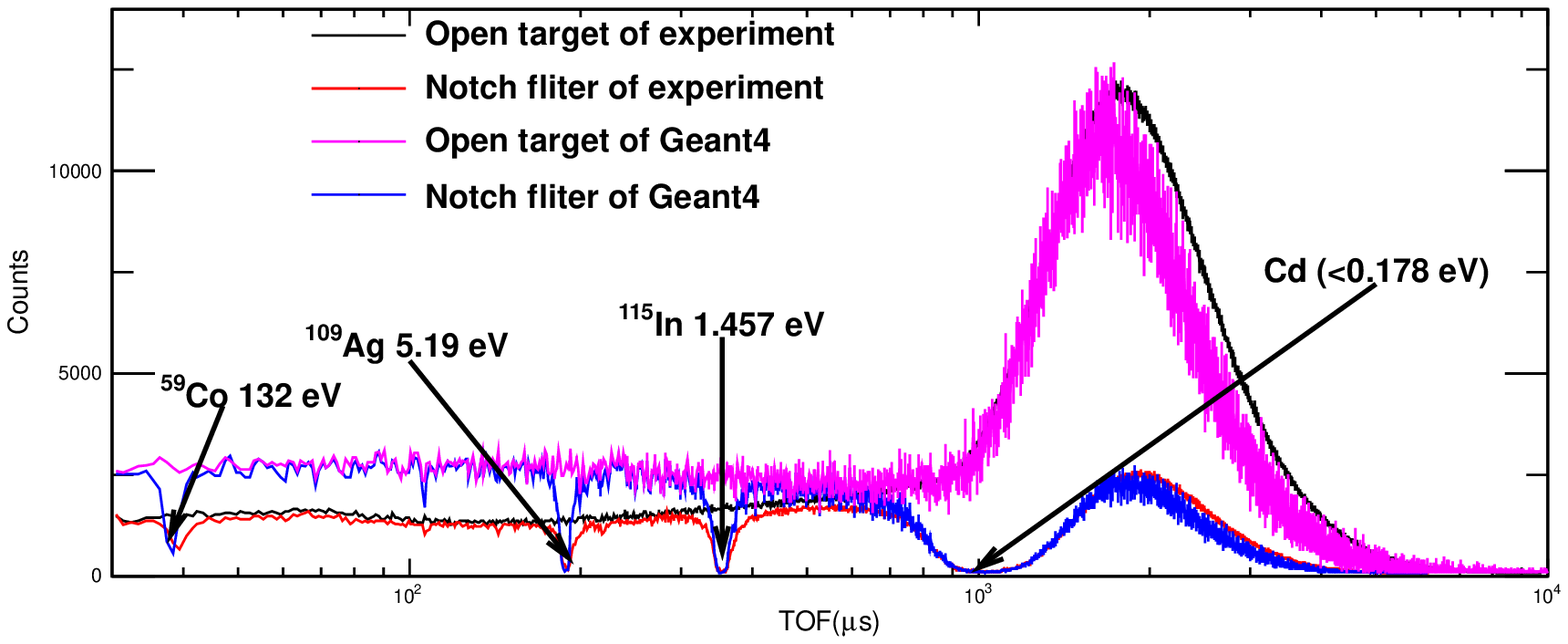}
\figcaption{\label{tofME}    TOF spectra for the open target and notch filter with the $B_{0}+B_{ne}(t)$ background
subtracted of the experiment compared with those obtained by the simulation. }
\end{center}

\subsection{Total Cross Section}

The neutron total cross section is determined by measuring the neutron beam transmitted through a known amount of sample
 and comparing it with the beam transmitted without sample.
The transmission rate of neutrons at the $i$th group energy $E_{i}$ is defined as the fraction of incident neutrons passing through
the sample compared to that in the open beam, which is related to the neutron total cross section $\sigma_{T}(E_{i})$ as follows:
\begin{equation}
\label{transmit}
T(E_{i})=\frac{[S(E_{i})/M_{S}-B_{S}(E_{i})/M_{B_{S}}]}{[O(E_{i})/M_{O}-B_{O}(E_{i})/M_{B_{O}}]},
\end{equation}
\begin{equation}
\label{cross}
\sigma_{T}(E_{i})=-\frac{1}{N}\ln{T(E_{i})},
\end{equation}
where $S(E_{i})$ is the count of the sample with a thickness of $N$ atoms per barn,
$O(E_{i})$ is the blank sample count,
$B_{S}(E_{i})$ is the background count of the sample,
and $B_{O}(E_{i})$ is the background count of the blank sample.
$M_{S}$, $M_{O}$, $M_{B_{S}}$ and $M_{B_{O}}$ are the normalized constants according to the monitored counts of the three samples.
$S(E_{i})$, $O(E_{i})$, $B_{S}(E_{i})$ and $B_{O}(E_{i})$ are all normalized to the counts of $O(E_{i})$ through the three normalized constants.

The resonance times of Co, Ag, and In were determined by fitting the absorption
peak as a function of TOF  by using the following fitting function,
\begin{equation}
\label{peak}
y=C_{0}+C_{1}t-(2A/\pi)w/[4(t-t_{c})^{2}+w^{2}],
\end{equation}
where $C_{0}$ is the starting neutron count, $C_{1}$ is the slope, $A$ is the
area of the absorption peak, $w$ is the width of the peak, and $t_{c}$ is the
position of the peak, i.e. the resonance time.
All the results are summarized in Table~\ref{time}.

\begin{center}
\tabcaption{ \label{time}  Resonance time of the corresponding neutron energy.}
\footnotesize
\begin{tabular*}{100mm}{c@{\extracolsep{\fill}}ccc}
\toprule Isotope & Neutron energy/eV & Experiment/$\mu$s   &  Simulation /$\mu$s \\
\hline
\hphantom{0}$^{59}$Co & 132.\hphantom{000} & \hphantom{0}39.50 & \hphantom{0}37.22 \\
$^{109}$Ag & \hphantom{00}5.19\hphantom{0} & 189.72 & 188.14 \\
$^{115}$In & \hphantom{00}1.457 & 355.23 & 353.29 \\
\bottomrule
\end{tabular*}
\end{center}

The flight path length $L$ was obtained from the resonance energy $E_{n}$ in electronvolts corresponding
to the channel number $I$ of the resonance time as indicated in Table~\ref{time}
 by using the following fitting function,
\begin{equation}
\label{energy}
E_{n}[eV]=(72.3\times{L}[m]/(I\times\Delta{W}-\tau_{0})[\mu{s}])^{2},
\end{equation}
where $\Delta{W}$ is the channel width of the time digitizer that we used 4 ns
 and $\tau_{0}$ is the time difference between  the initial time from the RF trigger and
 the real time zero when the neutron burst was produced.
  $L$ and $\tau_{0}$ of both experiment and simulation are shown in Table~\ref{cali}.
 $L$ of the simulation agreed well with that of the experiment with a little difference.
 The electron source in the simulation was set at 8 cm from the front surface of the target,
 so $\tau_{0}$ of the simulation was close to zero.
 The RF trigger of the experiment was delayed.

\begin{center}
\tabcaption{ \label{cali}  Flight path length and time difference.}
\footnotesize
\begin{tabular*}{80mm}{c@{\extracolsep{\fill}}ccc}
\toprule  & Length/m   &  $\tau_{0}$ /$\mu$s \\
\hline
Experiment & 5.903$\pm$0.004 & \hphantom{0}2.36$\pm$0.03 \\
Simulation & 5.928$\pm$0.010 & -0.09$\pm$0.06 \\
\bottomrule
\end{tabular*}
\end{center}

The total cross sections of beryllium were obtained in the neutron energy range from 0.007 to 0.1 eV by using
Eqs.~\ref{transmit} and ~\ref{cross}, as shown in Fig.\ref{harveyfold}.
The statistical uncertainty can be determined from Eq.~\ref{cross} if we assume that the monitor counters are equal
 during the measurements as follows:
\begin{equation}
\label{stat}
(\Delta\sigma_{stat.})_{i}=\frac{1}{N}\sqrt{\frac{S(E_{i})/M_{S}+B_{S}(E_{i})/M_{B_{S}}}{[S(E_{i})/M_{S}-B_{S}(E_{i})/M_{B_{S}}]^{2}}
+\frac{O(E_{i})/M_{O}+B_{O}(E_{i})/M_{B_{O}}}{[O(E_{i})/M_{O}-B_{O}(E_{i})/M_{B_{O}}]^{2}}}
,
\end{equation}
where subscript $i$ corresponds to each energy group, as in reference~\cite{Taofeng}.
The overall statistical uncertainties for the measured total cross sections are varied from $2\%$ to $6\%$,
depending on the neutron energy.

The present measurement is shown to be in excellent agreement with the fold high-accuracy measurement performed by
Harvey et al.~\cite{Harvey} with the response function of PNS in the energy range from 0.007 to 0.1 eV,
 as shown in Fig.\ref{harveyfold}.
It demonstrates that the uncertainties of the measurements at PNS are primarily due to
the broadening in time by the neutron transport in the target-moderator assembly.

\begin{center}
\includegraphics[width=15cm]{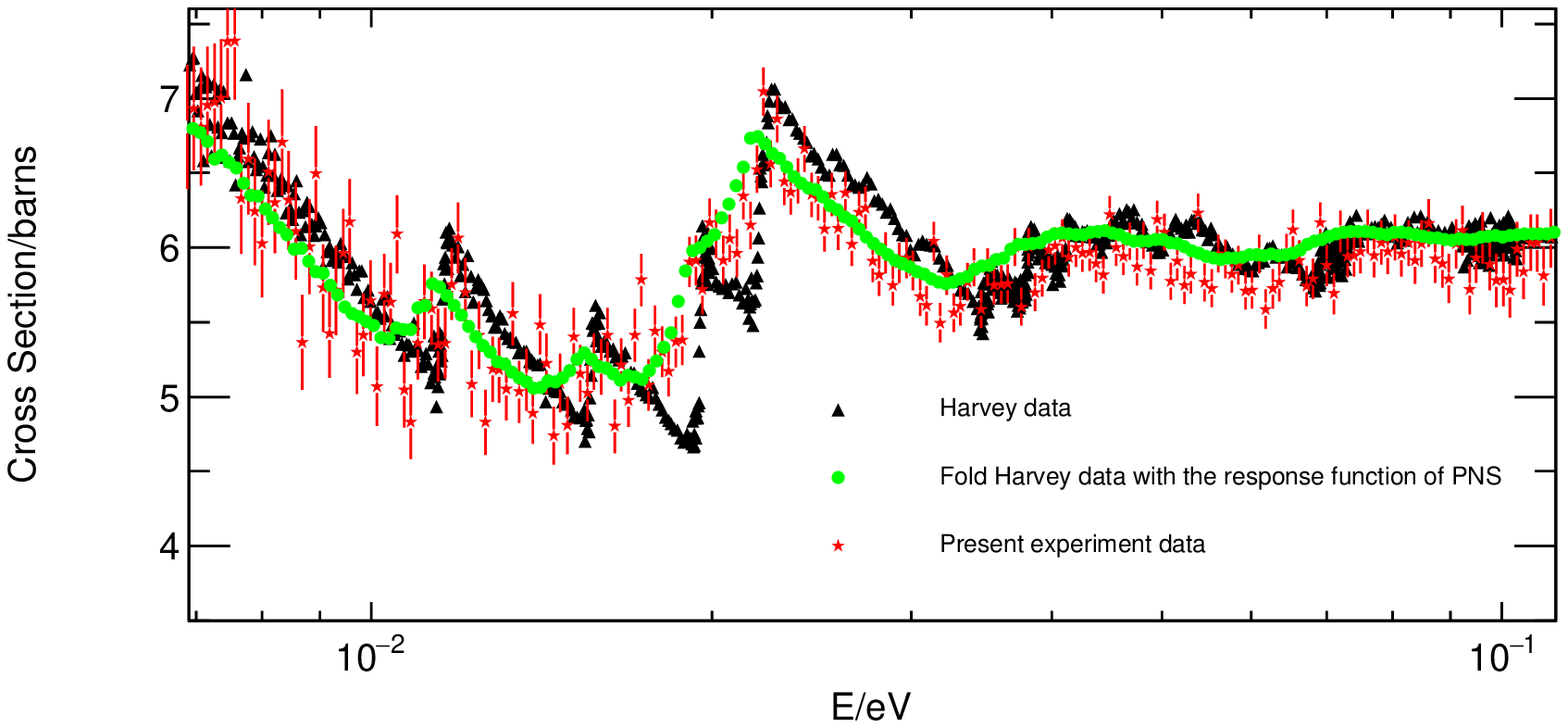}
\figcaption{\label{harveyfold}   Neutron total cross section of beryllium compared with the Harvey data
and the fold Harvey data with the response function of PNS. }
\end{center}

\section{Conclusion}

A system for measuring the neutron total cross section for thermal neutrons was described in detail in this paper.
The digital-signal-processing technique and a background measurement method were used in the system.
The energy resolutions due to the target-moderator assembly were obtained by using a Geant4 simulation.
In order to decrease the calculation time, the local weighted method was used in the simulation.
The neutron total cross section for beryllium was measured by this system in the range from 0.007 to 0.1 eV
and compared with that of the fold Harvey data.
The statistical uncertainties are varied from $2\%$ to $6\%$ depending on the neutron energy.

\paragraph{Acknowledgements}
\paragraph{}
The authors would like to express their sincere thanks to the free electron group and to other faculty members, and appreciate
their encouragement. The authors are grateful to P.Schillebeeckx of Joint Research Centre for suggestions and assistance in data analysis.
This work was supported by the Chinese TMSR Strategic Pioneer Science and Technology Project (No.XDA02010100), and
 National Natural Science Foundation of China (NSFC)(No.11475245,No.11305239).

\vspace{1mm}
\centerline{\rule{80mm}{0.1pt}}

\clearpage
\end{CJK*}
\end{document}